\documentclass[a4paper,12pt]{article}
\usepackage{epsfig}
\usepackage{color}
\usepackage{cite}

\renewcommand{\baselinestretch}{1.5}
\newlength{\dinwidth}
\newlength{\dinmargin}
\begin{document}
\newcommand{\beq}{\begin{eqnarray}}
\newcommand{\eeq}{\end{eqnarray}}
%%%%%%%%%%%%%%%%%%%%%%%%%%%%%%%%%%%%%%%%%%%%%%%%%

\title{Probing the R-parity violating supersymmetric effects in the exclusive $b \to c \ell^- \bar{\nu}_\ell$  decays}
\author{\small Jie Zhu$^{1,2}$,~Hua-Min Gan$^{1}$,~Ru-Min Wang$^{1}$\thanks{E-mail:ruminwang@sina.com},~Ying-Ying Fan$^{1}$,~Qin Chang$^{2}$,~Yuan-Guo Xu$^{1}$\thanks{E-mail:yuanguoxv@163.com}
  \\
  \renewcommand{\baselinestretch}{1.1}
{\scriptsize {$^1$ \it Institute of
    Theoretical Physics, Xinyang Normal University, Xinyang, Henan 464000, China
}}
\\
{\scriptsize  {$^2$ \it Institute of Particle and Nuclear Physics, Henan Normal University, Xinxiang,  Henan 453007, China
}}
}\renewcommand{\baselinestretch}{1.4}
 \maketitle
\vspace{-0.6cm}

Motivated by recent results from the LHCb, BABAR and Belle collaborations on $B\to D^{(*)}\ell^-\bar{\nu}_\ell$ decays, which significantly  deviate from the Standard Model  and hint the possible new physics  beyond the Standard Model, we probe the R-parity violating supersymmetric effects in $B^-_c\to \ell^-\bar{\nu}_\ell$ and $B\to D^{(*)}\ell^-\bar{\nu}_\ell$  decays.
We find that (i) $\mathcal{B}(B^-_c \to e^- \bar{\nu}_e)$ and $\mathcal{B}(B^-_c \to \mu^- \bar{\nu}_\mu)$ are sensitive to the constrained  slepton exchange couplings; (ii) the normalized forward-backward  asymmetries of $B \to D e^- \bar{\nu}_{e}$ decays have been greatly affected by the constrained  slepton exchange couplings, and  their signs could be changed; (iii) all relevant observables in the exclusive $b\to c\tau^- \bar{\nu}_\tau$ decays and the ratios $\mathcal{R}(D^{(*)})$ are sensitive to the slepton exchange coupling,
$\mathcal{R}(D^*)$ could be enhanced by the constrained slepton exchange coupling to reach each 95\% confidence level experimental range from BABAR, Belle and LHCb,  but it could not reach the lower limit of the 95\% confidence level experimental average.
Our results in this work could be used to probe R-parity violating effects, and will correlate with searches for direct supersymmetric signals at the running LHCb and the forthcoming Belle-II.

\vspace{1cm}{PACS Numbers: 13.20.He, 11.30.Fs,  12.60.Jv}
\vspace{1.5cm}
\section{Introduction}

The semileptonic decays $ B \to D^{(*)} \ell^- \bar{\nu}_\ell$ are very important processes in testing the Stand Model (SM) and in searching for the new physics (NP) beyond the SM, for example, the extraction of the Cabbibo-Kobayashi-Maskawa matrix element $|V_{cb}|$.
The semileptonic decays $B\to D^{(*)} \ell^- \bar{\nu}_\ell$ have been measured by the CLEO\cite{Adam:2002uw}, Belle\cite{Dungel:2010uk,Huschle:2015rga}, BABAR\cite{Aubert:2007rs,Aubert:2008yv,Lees:2012xj,Lees:2013uzd} and LHCb \cite{Aaij:2015yra} collaborations. For the ratios $\mathcal{R}(D^{(*)})\equiv \frac{\mathcal{B}(B\to D^{(*)}\tau^-\bar{\nu}_\tau)}{\mathcal{B}(B\to D^{(*)}\ell'^-\bar{\nu}_{\ell'})}$ with $\ell'=e$ or $\mu$,  the experimental averages from  the Heavy Flavor Averaging Group \cite{Amhis:2014hma} are
\beq
\label{eq:exp01}
{\cal R}(D)^{\rm Exp.} = 0.391 \pm 0.050,\nonumber\\
{\cal R}(D^*)^{\rm Exp.} = 0.322 \pm 0.021,
\eeq
the SM predictions  \cite{Fajfer:2012vx,Kamenik:2008tj} are
\beq
\label{eq:exp01}
{\cal R}(D)^{\rm SM} = 0.297 \pm 0.017,\nonumber\\
{\cal R}(D^*)^{\rm SM} = 0.252 \pm 0.003,
\eeq
the experimental measurements of $\mathcal{R}(D)$ and  $\mathcal{R}(D^*)$ differ from their SM predictions by  1.7$\sigma$  and 3.0 $\sigma$ deviations, respectively,  and these hint the possible NP beyond the SM.

The exclusive $b \to c\ell^- \bar{\nu}_\ell$ decays have been studied extensively in the framework of the SM and various NP models, for instance,  see Refs. \cite{lattice-Bailey,lattice-Na,Bailey:2012jg,Bailey:2012rr,Faustov:2012nk,Fajfer:2012jt,Datta:2012qk,
Choudhury:2012hn,Bhattacharya:2014wla,Becirevic:2012jf,He:2012zp,Fajfer:2013aw,Pich:2013kg,Ricciardi:2013jf,
Fan:2013qz,Xiao:2014ana,Fan:2015kna,Stone:2012yr,Sakaki:2012ft,Crivellin:2012ye,Celis:2013jha,Celis:2012dk,Becirevic:2016hea}.
The R-parity violating (RPV) supersymmetry (SUSY) is one of the respectable NP models that survived electroweak data~\cite{Chemtob:2004xr,Nandi:2006qe,Xu:2006vk,Kim:2007uq,Kim:2009mp,Wang:2010vv,Wang:2011aa,Cheng:2013qpa,Dreiner:2013jta}. In this paper, we will explore the RPV effects in the leptonic and semileptonic  exclusive $b \to c \ell^- \bar{\nu}_\ell$  decays.
We constrain  relevant RPV parameter spaces from present experimental measurements and analyze their contributions to  the branching ratios, the differential branching ratios, the normalized forward-backward (FB) asymmetries of the charged leptons, and the ratios of the branching ratios of relevant semileptonic $B$ decays.

The paper is organized as follows. In section 2, we briefly review the theoretical results
of the exclusive $b \to c \ell^- \bar{\nu}_\ell$
decays in the RPV SUSY model. In section 3, using the constrained parameter spaces from relevant experimental measurements, we make a detailed classification research on the RPV effects on the quantities which have not been measured or not been well measured yet. Our conclusions are given in section 4.

\section{The exclusive $b \to c \ell^- \bar{\nu}_{\ell}$  decays in the SUSY without R-parity}

In the RPV SUSY model, the similar processes $b \to u \ell^- \bar{\nu}_\ell$ have been studied in Ref.  \cite{Kim:2007uq}, and we will only give the final expressions in this section.

The branching ratio for the pure leptonic decays $B_c^- \to \ell^-_m \bar{\nu}_{\ell_n}$ can be written as \cite{Kim:2007uq}
\begin{eqnarray}{\small
\mathcal{B}(B_c^- \to \ell^-_m \bar{\nu}_{\ell_n})=\left| \frac{G_F}{\sqrt{2}}V_{cb} - \sum_i \frac{\lambda'_{n3i}\tilde{\lambda}'^*_{m2i}}{8m^2_{\tilde{d}_{iR}}}+\sum_i \frac{\lambda_{inm}\tilde{\lambda}'^*_{i23}}{4m^2_{\tilde{\ell}_{iL}}}\frac{\mu_{B_c}}{m_{\ell}}\right|^2
\frac{ \tau_{B_c}}{4 \pi}f_{B_c}^2 m_{B_c} m_{\ell}^2\left[1-\frac{m_{\ell}^2}{m_{B_c}^2} \right]^2,}
\end{eqnarray}
where $\mu_{B_c}\equiv m^2_{B_c}/(\bar{m}_b+\bar{m}_c)$.

The differential branching ratios for the semileptonic decays $ B \to D \ell_m^- \bar{\nu}_{\ell_n}$ could be written as \cite{Kim:2007uq}
\begin{eqnarray}
\frac{d\mathcal{B}(B \to D \ell^- \bar{\nu}_{\ell_n})}{ds d\rm{cos}\theta}=\frac{\tau_B\sqrt{\lambda_D}}{2^7\pi^3m^3_B}\left(1-\frac{m^2_{\ell_m}}{s}\right)^2\left[N^D_0+N^D_1 \rm{cos}\theta+N^D_2\rm{cos}^2\theta\right],
\end{eqnarray}
with
\begin{eqnarray}
N^D_0&=&\left|\frac{G_F}{\sqrt{2}}V_{cb}-\sum_i\frac{\lambda'_{n3i}\tilde{\lambda}'^*_{m2i}}{8m^2_{\tilde{d}_{iR}}}\right|^2[f^D_+(s)]^2\lambda_D
+\left|\frac{G_F}{\sqrt{2}}V_{cb}-\sum_i\frac{\lambda'_{n3i}\tilde{\lambda}'^*_{m2i}}{8m^2_{\tilde{d}_{iR}}} \right.\nonumber\\
&&\left.+\sum_i\frac{\lambda_{inm}\tilde{\lambda}'^*_{i23}}{4m^2_{\tilde{\ell}_{iL}}}\frac{s}{m_{\ell_m}(\bar{m}_b-\bar{m}_c)}\right|^2
m^2_{\ell_m}[f^D_0(s)]^2\frac{(m^2_B-m^2_D)^2}{s},\\
N^D_1&=&\left\{\left|\frac{G_F}{\sqrt{2}}V_{cb}-\sum_i\frac{\lambda'_{n3i}\tilde{\lambda}'^*_{m2i}}{8m^2_{\tilde{d}_{iR}}}\right|^2
+Re\left[\left(\frac{G_F}{\sqrt{2}}V_{cb}-\sum_i\frac{\lambda'_{n3i}\tilde{\lambda}'^*_{m2i}}{8m^2_{\tilde{d}_{iR}}}\right)^{\dagger}\right.\right.\nonumber\\
&&\left.\left.\times\sum_i\frac{\lambda_{inm}\tilde{\lambda}'^*_{i23}}{4m^2_{\tilde{\ell}_{iL}}}\frac{s}{m_{\ell_m}(\bar{m}_b-\bar{m}_c)}\right]\right\}
2m^2_{\ell_m}f^D_0(s)f^D_+(s)\sqrt{\lambda_D}\frac{(m^2_B-m^2_D)}{s},\\
N^D_2&=&-\left|\frac{G_F}{\sqrt{2}}V_{cb}-\sum_i\frac{\lambda'_{n3i}\tilde{\lambda}'^*_{m2i}}{8m^2_{\tilde{d}_{iR}}}\right|^2[f^D_+(s)]^2\lambda_D\left(1-\frac{m^2_{\ell_m}}{s}\right),
\end{eqnarray}

The differential branching ratios for the semileptonic decays $ B \to D^* \ell_m^- \bar{\nu}_{\ell_n}$ could be written as
\cite{Kim:2007uq}
\begin{eqnarray}
\frac{d\mathcal{B}(B \to D^* \ell_m^- \bar{\nu}_{\ell_n})}{ds d\rm{cos}\theta}=\frac{\tau_B\sqrt{\lambda_D^*}}
{2^7\pi^3m^3_B}\left(1-\frac{m^2_{\ell_m}}{s}\right)^2\left[N^{D^*}_0+N^{D^*}_1 \rm{cos}\theta+N^{D^*}_2\rm{cos}^2\theta\right],
\end{eqnarray}
with
\begin{eqnarray}
N^{D^*}_0&=&\left|\frac{G_F}{\sqrt{2}}V_{cb}-\sum_i\frac{\lambda'_{n3i}\tilde{\lambda}'^*_{m2i}}{8m^2_{\tilde{d}_{iR}}}\right|^2
\left\{[A^{D^*}_1(s)]^2\left(\frac{\lambda_{D^*}}{4m^2_{D^*}}+(m^2_{\ell_m}+2s)\right)(m_B+m_{D^*})^2\right.\nonumber\\
&&+[A^{D^*}_2(s)]^2\frac{\lambda_{D^*}^2}{4m_V^2(m_B+m_{D^*})^2}+[V^{D^*}(s)]^2\frac{\lambda_{D^*}}{(m_B+m_{D^*})^2}(m_{\ell_m}^2+s)\nonumber\\
&&\left.-A_1^{D^*}(s)A_2^{D^*}(s)\frac{\lambda_{D^*}}{2m^2_{D^*}}(m^2_B-s-m_{D^*}^2)\right\}
+\left|\frac{G_F}{\sqrt{2}}V_{cb}\right.\nonumber\\
&&\left.-\sum_i\frac{\lambda'_{n3i}\tilde{\lambda}'^*_{m2i}}{8m^2_{\tilde{d}_{iR}}}+
\sum_i\frac{\lambda_{inm}\tilde{\lambda}'^*_{i23}}{4m^2_{\tilde{\ell}_{iL}}}\frac{s}{m_{\ell_m}(\bar{m}_b+\bar{m}_c)}\right|^2
[A^{D^*}_0(s)]^2\frac{m^2_{\ell_m}}{s}\lambda_{D^*},\\
N^{D^*}_1&=&\left\{\left|\frac{G_F}{\sqrt{2}}V_{cb}-\sum_i\frac{\lambda'_{n3i}\tilde{\lambda}'^*_{m2i}}{8m^2_{\tilde{d}_{iR}}}\right|^2+Re\left[\left(\frac{G_F}{\sqrt{2}}V_{cb}-\sum_i\frac{\lambda'_{n3i}\tilde{\lambda}'^*_{m2i}}{8m^2_{\tilde{d}_{iR}}}\right)^{\dagger}\right.\right.\nonumber\\
&&\left.\left.\sum_i\frac{\lambda_{inm}\tilde{\lambda}'^*_{i23}}{4m^2_{\tilde{\ell}_{iL}}}\frac{s}{m_{\ell_m}(\bar{m}_b+\bar{m}_c)}\right]\right\}\nonumber\\
&&\times\left[A^{D^*}_0(s)A^{D^*}_1(s)\frac{m_{\ell_m}^2(m_B+m_{D^*})(m^2_B-m^2_{D^*}-s)\sqrt{\lambda_{D^*}}}{sm_{D^*}}\right.\nonumber\\
&&\left.-A^{D^*}_0(s)A^{D^*}_2(s)\frac{m^2_{\ell_m}\lambda^{3/2}_{D^*}}{sm_{D^*}(m_B+m_{D^*})}\right]\nonumber\\
&&+\left|\frac{G_F}{\sqrt{2}}V_{cb}-\sum_i\frac{\lambda'_{n3i}\tilde{\lambda}'^*_{m2i}}{8m^2_{\tilde{d}_{iR}}}\right|^2
A^{D^*}_1(s)V^{D^*}(s)4s\sqrt{\lambda_{D^*}},\\
N^{D^*}_2&\!=\!&-\left|\frac{G_F}{\sqrt{2}}V_{cb}-\sum_i\frac{\lambda'_{n3i}\tilde{\lambda}'^*_{m2i}}{8m^2_{\tilde{d}_{iR}}}\right|^2\!\left(1-\frac{m^2_{\ell_m}}{s}\right)
\!\lambda_{D^*}\!\left\{[A^{D^*}_1(s)]^2\frac{(m_B+m_{D^*})^2}{4m_{D^*}^2}\right.\nonumber\\
&&+[V^{D^*}(s)]^2\frac{s}{(m_B+m_{D^*})^2}+[A^{D^*}_2(s)]^2\frac{\lambda_{D^*}}{4m_{D^*}^2(m_B+m_{D^*})^2}\nonumber\\
&&\left.-A^{D^*}_1(s)A^{D^*}_2(s)\frac{m_B^2-m_{D^*}^2-s}{2m_{D^*}^2}\right\},
\end{eqnarray}
where $s=q^2=(p_B-p_{D^{(*)}})^2$, the kinematic factor
$\lambda_{D^{(*)}}=m^4_B+m^4_{D^{(*)}}+s^2-2m^2_Bm^2_{D^{(*)}}-2m^2_Bs-2m^2_{D^{(*)}}s$, and the $\theta$ is the angle between the momentum of $B$ meson and the charged lepton in the c.m. system of $\ell-\nu$.

The normalized forward-backward asymmetry of the charged lepton $\bar{\mathcal{A}}_{FB}^{D^{(*)}}$ are given as~\cite{Kim:2007uq}
\begin{eqnarray}
\bar{\mathcal{A}}_{FB}(B\to D^{(*)}\ell^-_m \bar{\nu}_{\ell_n})=\frac{N^{D^{(*)}}_1}{2N^{D^{(*)}}_0+2/3N^{D^{(*)}}_2}.
\end{eqnarray}
From above expressions, we can see that, unlike the contributions of the squark exchange couplings $\lambda'_{n3i}\tilde{\lambda}'^*_{m2i}$ and the SM contributions, the slepton exchange couplings $\lambda_{inm}\tilde{\lambda}'^*_{i23}$ will not be suppressed by $s$ and helicity.

\section{Numerical Results and Discussions}

In the numerical calculations, the main theoretical input parameters are the transition form factors, the decay constant of $B^-_c$ meson, the masses, the mean lives, the CKM matrix element, etc. For the transition form factors,
the traditional approaches to calculate the relevant transition form factors are the heavy quark effective theory ~\cite{Fajfer:2012vx,Fajfer:2013aw}, the Lattice QCD techniques~\cite{lattice-Bailey,lattice-Na} and the pQCD factorization approach with and without Lattice QCD input \cite{Fan:2013qz,Xiao:2014ana,Fan:2015kna}, we will  use the form factors based on the heavy quark effective theory~\cite{Fajfer:2012vx,Fajfer:2013aw}.
The decay constant of $B^-_c$ meson is  taken  from Ref. \cite{Colquhoun:2015oha}, and the rest of the theoretical input parameters  are taken from the  Particle Data Group (PDG)~\cite{Agashe:2014kda}.
Notice that  we assume the masses of the corresponding slepton are 500 GeV.
For other values of
the slepton masses, the bounds on the couplings in this
paper can be easily obtained by scaling them by factor of
$\tilde{f}^2\equiv\left(\frac{m_{\tilde{\ell}}}{500\rm{GeV}}\right)^2$.

In our calculation, we consider only one NP coupling at one time and keep its interference with the SM amplitude to study the RPV SUSY effects. To be conservative, the input parameters and the experimental bounds except for $\mathcal{B}(B \to D^* \tau^- \bar{\nu}_{\tau})$ and  $\mathcal{R}(D^*)$ at $95\%$ confidence level (CL) will be used to constrain parameter spaces of the relevant new couplings. Noted that we do not impose the experimental bounds
from $\mathcal{B}(B \to D^* \tau^- \bar{\nu}_{\tau})$ and $\mathcal{R}(D^*)$, since their experimental measurements  obviously deviate from their SM predictions, and we leave them as predictions
of the restricted parameter spaces of the RPV couplings,
and then compare them with the experimental results.

Due to the strong helicity suppression, the squark exchange couplings have no very obvious effects on the differential branching ratios and the normalized FB asymmetries of the semileptonic  exclusive $b \to c \ell^-_m \bar{\nu}_{\ell_n}$  decays. So we will only focus on the slepton exchange couplings in our follow discussions.
For the slepton exchange couplings, $\lambda_{i11}\tilde{\lambda}'^*_{i23}$  and  $\lambda_{i22}\tilde{\lambda}'^*_{i23}$ , which contribute to both $b\to c \ell'^-_m \bar{\nu}_{\ell'_n}$ and $b\to s \ell'^+_m \ell'^-_n$ transitions, the stronger constraints are from the exclusive  $b\to s \ell'^+_m \ell'^-_n$ decays~\cite{Xu:2006vk,Wang:2011aa},  % $|\lambda_{i11}\tilde{\lambda}'^*_{i23}|<5.75\times10^{-4}$ and $|\lambda_{i22}\tilde{\lambda}'^*_{i23}|<1.63\times10^{-5}$,
 nevertheless, the RPV weak phases of the two slepton exchange couplings  are not obviously constrained by current experimental measurements.

\subsection{The exclusive $b\to c e^- \bar{\nu}_e$ decays}
First, we focus on slepton exchange  couplings $\lambda_{i11}\tilde{\lambda}'^*_{i23}$ contribute to five decay modes, $B^-_c \to e^- \bar{\nu}_e$, $B^-_u \to D^0_u    e^- \bar{\nu}_e$, $B^-_u \to D^{*0}_u e^- \bar{\nu}_e$, $B^0_d \to D^+_d    e^- \bar{\nu}_e$ and $B^0_d \to D^{*+}_d e^- \bar{\nu}_e$ decays.
The branching ratios of four semileptonic processes have been accurately measured by CLEO~\cite{Adam:2002uw}, Belle \cite{Dungel:2010uk} and BABAR \cite{Aubert:2008yv,Aubert:2007rs} collaborations.
The  95\% CL ranges of the experimental average values from the PDG~\cite{Agashe:2014kda} are listed in the second column of Table~\ref{tab:bcelslp}.
The SM predictions at 95\% CL are presented in the third column  of Table~\ref{tab:bcelslp}.

\begin{table}[t]
\caption{Branching ratios of the exclusive $b\to c e^-\bar{\nu}_e $ decays (in units of $10^{-2}$) except for $\mathcal{B}(B^-_c \to e^- \bar{\nu}_e)$ (in units of $10^{-9}$). The experimental ranges and the SM predictions at 95\% CL  are listed in the second and third columns, respectively. In the last column, ``SUSY/$\lambda_{i11}\tilde{\lambda}'^*_{i23}$" denotes the SUSY predictions considering the constrained $\lambda_{i11}\tilde{\lambda}'^*_{i23}$ couplings. The similar in Table~\ref{tab:bcmulslp} and Table~\ref{tab:bctaulslp}.}
\begin{center}
\begin{tabular}{lccc}
\hline\hline
 Observable                                         & Exp. data       & SM predictions & SUSY/$\lambda_{i11}\tilde{\lambda}'^*_{i23}$  \\\hline
$\mathcal{B}(B^-_c \to          e^- \bar{\nu}_e)$   &$\cdots$         &$[1.39,~2.72]$     &$[1.49\times10^{-2},1068]$            \\\hline
$\mathcal{B}(B^-_u \to D^0_u    e^- \bar{\nu}_e)$   &$[2.05,~2.49]$   &$[1.81,~2.91]$     &$[2.10,2.49]$                   \\\hline
$\mathcal{B}(B^-_u \to D^{*0}_u e^- \bar{\nu}_e)$   &$[5.32,~6.06]$   &$[4.78,~5.81]$     &$[5.34,5.61]$                   \\\hline
$\mathcal{B}(B^0_d \to D^+_d    e^- \bar{\nu}_e)$   &$[1.95,~2.43]$   &$[1.68,~2.69]$     &$[1.96,2.33]$                   \\\hline
$\mathcal{B}(B^0_d \to D^{*+}_d e^- \bar{\nu}_e)$   &$[4.71,~5.15]$   &$[4.44,~5.38]$     &$[4.89,5.15]$                   \\\hline
\hline
\end{tabular}
\end{center}
\label{tab:bcelslp}
\end{table}

Using the experimental bounds of  relevant exclusive $b\to c \ell^- \bar{\nu}_\ell$ decays at 95\% CL, we obtain the slepton exchange couplings $|\lambda_{i11}\tilde{\lambda}'^*_{i23}|\leq0.22$. At present, the strongest bounds on the slepton exchange couplings come from the exclusive  $b\to s e^+e^-$ decays, $|\lambda_{i11}\tilde{\lambda}'^*_{i23}|\leq5.75\times10^{-4}$  with 500 GeV slepton masses \cite{Xu:2006vk}, which will be used in our numerical results.
In addition, the experimental bounds at the 95\% CL   listed in the second column of Table~\ref{tab:bcelslp} are also considered to further constrain the slepton exchange couplings.
Our numerical results of relevant branching ratios, which consider the constrained slepton exchange couplings,  are listed in the last column of Table \ref{tab:bcelslp}, and we can see that the constrained slepton exchange coupling has significant effects on $\mathcal{B}(B^-_c \to e^-\bar{\nu}_e)$, which could be suppressed 2 orders or enhanced  3 orders  by the constrained slepton exchange couplings. Nevertheless, the constrained slepton exchange couplings have no significant effects on the branching ratios of relevant semileptonic decays.

For $B^-_u \to D^{(*)0}_u  e^- \bar{\nu}_e$ and $B^0_d \to D^{(*)+}_d  e^- \bar{\nu}_e$ decays, since the $SU(3)$ flavor symmtry implies $\mathcal{M}(B^-_u \to D^{(*)0}_u  e^- \bar{\nu}_e)
    \simeq\mathcal{M}(B^0_d \to D^{(*)+}_d  e^- \bar{\nu}_e)$,  the slepton exchange RPV contributions to $B^-_u \to D^{(*)0}_u  e^- \bar{\nu}_e$
     and $B^0_d \to D^{(*)+}_d  e^- \bar{\nu}_e$  are very similar to each other. So we would take $B^-_u \to D^{(*)0}_u e^- \bar{\nu}_e$ decays as examples. The similar in the exclusive $b\to c\mu^- \bar{\nu}_\mu$ and $b\to c\tau^- \bar{\nu}_\tau $ decays.

\begin{figure}[t]
\begin{center}
\includegraphics[scale=0.7]{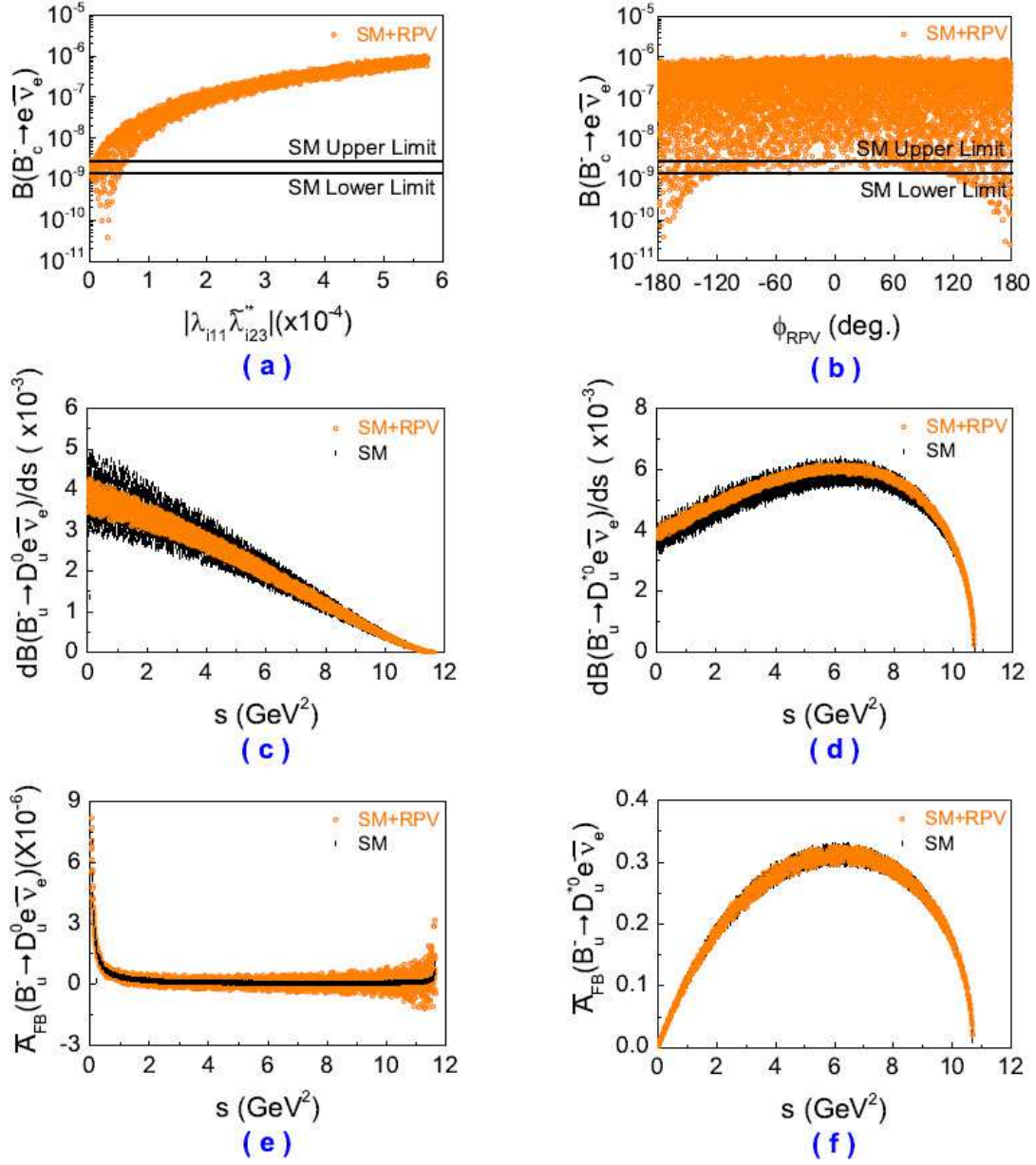}
\end{center}
 \caption{The constrained slepton exchange coupling effects in the exclusive $b \to c e^- \bar{\nu}_e$ decays. }
 \label{fig:bcelslp}
\end{figure}

Fig.~\ref{fig:bcelslp} shows the constrained RPV effects of $\lambda_{i11}\tilde{\lambda}'^*_{i23}$ on  $\mathcal{B}(B^-_c \to          e^- \bar{\nu}_e)$, $d\mathcal{B}(B_u^- \to D^{(*)0} e^- \bar{\nu}_e)/ds $, and  $\bar{\mathcal{A}}_{FB}(B_u^- \to D^{(*)0} e^- \bar{\nu}_e)$. The SM results are also displayed for comparing.
Comparing the RPV SUSY predictions to the SM ones, we have the following remarks.
\begin{itemize}
\item As shown in Fig. \ref{fig:bcelslp} (a-b), $\mathcal{B}(B^-_c \to e^- \bar{\nu}_e)$ is very sensitive to both moduli and weak phases of the $\lambda_{i11}\tilde{\lambda}'^*_{i23}$ couplings, and this is due to that the slepton exchange coupling  effects on  $\mathcal{B}(B^-_c \to e^- \bar{\nu}_e)$ is increased by $m_B/m_e$.

\item As displayed in Fig. \ref{fig:bcelslp} (c-d), there are no obvious RPV effect on $ d\mathcal{B}(B^-_u \to D^{(*)0}_u e^- \bar{\nu}_e)$, since the present accurate experimental measurements of $\mathcal{B}(B^-_u \to D^{*0}_u e^- \bar{\nu}_e, B^0_d \to D^{*+}_d    e^- \bar{\nu}_e)$ give very strongly constraints on the slepton exchange couplings. For the same reason,  the branching ratios of relevant semileptonic decays are not sensitive to both moduli and weak phases of the $\lambda_{i11}\tilde{\lambda}'^*_{i23}$ couplings, so we do not display them in Fig. \ref{fig:bcelslp}.

  \item Fig.~\ref{fig:bcelslp} (e) shows us that  the constrained $\lambda_{i11}\tilde{\lambda}'^*_{i23}$ couplings provide quite obvious effects on $\bar{\mathcal{A}}_{FB}(B^-_u \to D^0_u e^- \bar{\nu}_e)$, its sign could be changed, nevertheless, this quantity is  tiny. Fig.~\ref{fig:bcelslp} (f) shows that there is  no obvious RPV effect on $\bar{\mathcal{A}}_{FB}(B^-_u \to D^{*0}_u e^- \bar{\nu}_e)$.
\end{itemize}

\subsection{The exclusive $b\to c\mu^-\bar{\nu}_\mu$ decays}

Now we pay attention to the contributions of the slepton exchange  couplings $\lambda_{i22}\tilde{\lambda}'^*_{i23}$  to $B^-_c \to \mu^- \bar{\nu}_{\mu}$,  $B^-_u \to D^0_u\mu^- \bar{\nu}_{\mu}$, $B^-_u \to D^{*0}_u \mu^- \bar{\nu}_{\mu}$, $B^0_d \to D^+_d    \mu^- \bar{\nu}_{\mu}$, $B^0_d \to D^{*+}_d \mu^- \bar{\nu}_{\mu}$ decays.
The four semi-leptonic decay branching ratios have been accurately measured  by CLEO~\cite{Adam:2002uw}, Belle \cite{Dungel:2010uk} and BABAR \cite{Aubert:2008yv,Aubert:2007rs} collaborations.
The experimental average values and the SM predictions at 95\% CL  are listed in the second and third column of Table~\ref{tab:bcmulslp}, respectively.
 \begin{table}[b]
\caption{Branching ratios of the exclusive $b\to c\mu^-\bar{\nu}_\mu $ decays
(in units of $10^{-2}$) except for $\mathcal{B}(B^-_c \to \mu^- \bar{\nu}_\mu)$
(in units of $10^{-4}$). }
\begin{center}
\begin{tabular}{lcccc}
\hline\hline
 Observable                                              & Exp. data       & SM predictions    & SUSY/$\lambda^*_{i22}\tilde{\lambda}'^*_{i23}$  \\\hline
$\mathcal{B}(B^-_c \to          \mu^- \bar{\nu}_\mu)$    &$\cdots$         &$[0.59,~1.16]$      &$[0.51,~1.17]$       \\\hline
$\mathcal{B}(B^-_u \to D^0_u    \mu^- \bar{\nu}_\mu)$    &$[2.05,~2.49]$   &$[1.81,~2.89]$      &$[2.09,~2.48]$        \\\hline
$\mathcal{B}(B^-_u \to D^{*0}_u \mu^- \bar{\nu}_\mu)$    &$[5.32,~6.06]$   &$[4.76,~5.77]$      &$[5.32,~5.59]$        \\\hline
$\mathcal{B}(B^0_d \to D^+_d    \mu^- \bar{\nu}_\mu)$    &$[1.95,~2.43]$   &$[1.68,~2.67]$      &$[1.96,~2.32]$        \\\hline
$\mathcal{B}(B^0_d \to D^{*+}_d \mu^- \bar{\nu}_\mu)$    &$[4.71,~5.15]$   &$[4.42,~5.35]$      &$[4.87,~5.13]$        \\\hline
\hline
\end{tabular}
\end{center}
\label{tab:bcmulslp}
\end{table}
\begin{figure}[t]
\begin{center}
\includegraphics[scale=0.7]{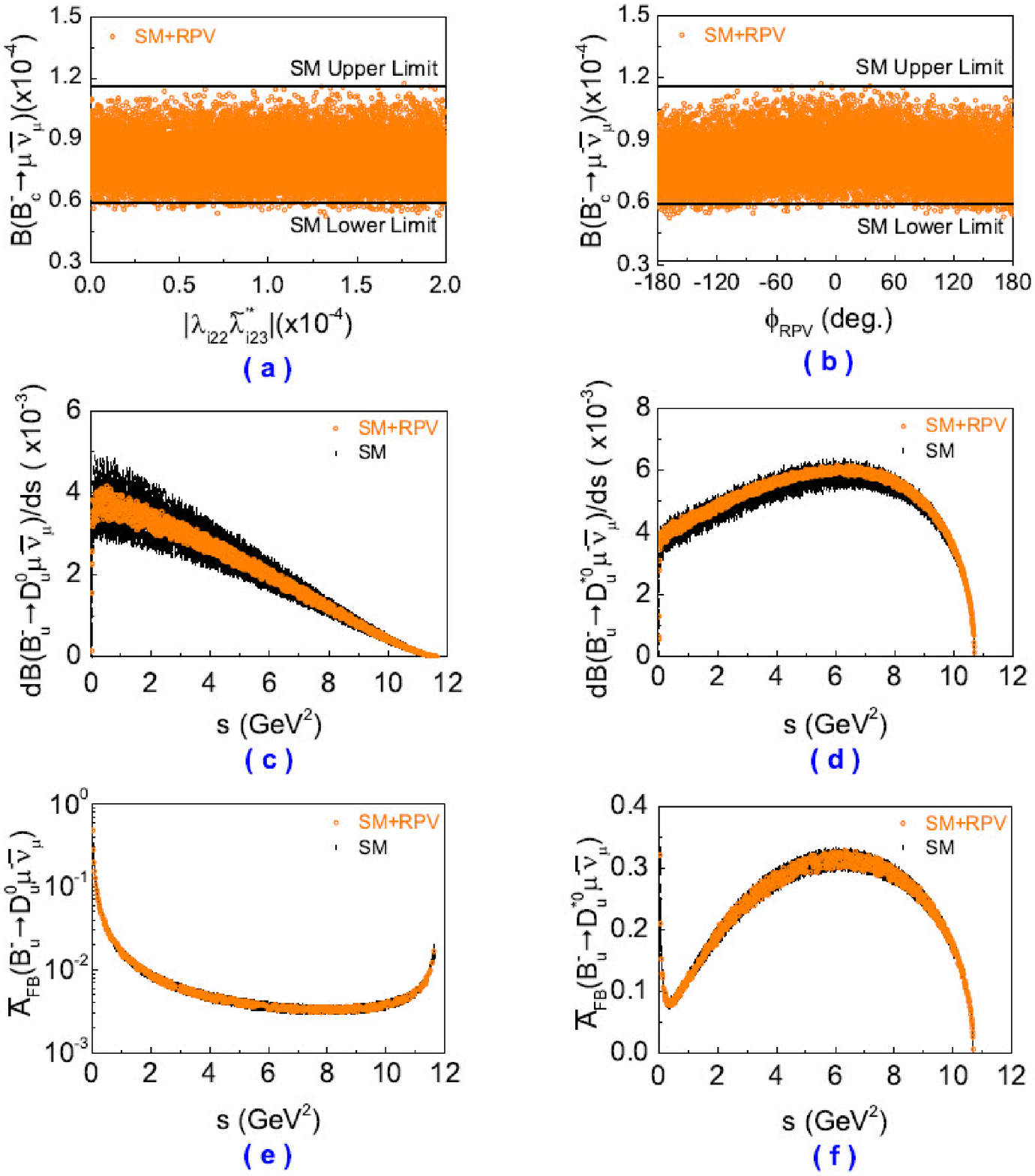}
\end{center}
 \caption{
 The constrained slepton exchange coupling effects in the exclusive  $b\to c \mu^- \bar{\nu}_{\mu}$  decays. }
 \label{fig:bcmulslp}
\end{figure}

We get the slepton exchange couplings $|\lambda_{i22}\tilde{\lambda}'^*_{i23}|<0.24$ from the exclusive $b\to c\ell^-\bar{\nu}_\ell$ decays, which are a lot  weaker than ones from the exclusive  $b\to s \mu^+ \mu^-$ decays, $|\lambda_{i22}\tilde{\lambda}'^*_{i23}|<2.0\times10^{-4}$ with 500 GeV slepton masses~\cite{Wang:2011aa}.
Taking the strongest bounds from the exclusive  $b\to s \mu^+ \mu^-$ decays and further considering the experimental bounds from the exclusive $b\to c\ell^-\bar{\nu}_\ell $ decays,   we predict the constrained slepton exchange effects  in the exclusive $b \to c \mu^- \bar{\nu}_{\mu}$ decays, which   are  given in  the last column of Table~\ref{tab:bcmulslp} and  displayed in Fig.~\ref{fig:bcmulslp}.
From  Table~\ref{tab:bcmulslp} and  Fig.~\ref{fig:bcmulslp}, we make the following points.

\begin{itemize}
  \item As displayed in Fig.~\ref{fig:bcmulslp} (a-b), $\mathcal{B}(B^-_c \to \mu^- \bar{\nu}_\mu)$ has some sensitivities  to both modulus and weak phases of the $\lambda_{i22}\tilde{\lambda}'^*_{i23}$ couplings, and it has maximum at $\phi_{RPV} \in [-60^{\circ}, 60^{\circ}]$.

\item  Fig.~\ref{fig:bcmulslp} (c-d) shows that the constrained slepton exchange couplings  have no obvious contribution to  $ d\mathcal{B}(B^-_u \to D^{(*)0}_u \mu^- \bar{\nu}_\mu)/ds$, and they are strongly constrained by present experimental data.

  \item As displayed in Fig.~\ref{fig:bcmulslp} (e-f), the constrained slepton exchange couplings also have no obvious contribution to $ \bar{\mathcal{A}}_{FB}(B^-_u \to D^{(*)0}_u \mu^- \bar{\nu}_\mu)$ at all $s$ range. Noted that, the slepton exchange coupling effects on  $\bar{\mathcal{A}}_{FB}(B^-_u \to D^0_u \mu^- \bar{\nu}_\mu)$ are very different from ones on $\bar{\mathcal{A}}_{FB}(B^-_u \to D^0_u e^- \bar{\nu}_e)$ displayed in Fig.~\ref{fig:bcelslp} (e-f), since the bounds on $|\lambda_{i22}\tilde{\lambda}'^*_{i23}|$  are about 3 times smaller than ones on $|\lambda_{i11}\tilde{\lambda}'^*_{i23}|$ (the same order of magnitude) and $\bar{\mathcal{A}}_{FB}(B^-_u \to D^0_u \mu^- \bar{\nu}_\mu)$ is 1000 times larger than  $\bar{\mathcal{A}}_{FB}(B^-_u \to D^0_u e^- \bar{\nu}_e)$.

\end{itemize}

\subsection{The exclusive $b\to c\tau^- \bar{\nu}_\tau$ decays}

In this subsection, we concentrate on the contributions of the slepton exchange couplings $\lambda_{i33}\tilde{\lambda}'^*_{i23}$  in  $B^-_c \to \tau^- \bar{\nu}_{\tau}$, $B^-_u \to D^0_u    \tau^- \bar{\nu}_{\tau}$, $B^-_u \to D^{*0}_u \tau^- \bar{\nu}_{\tau}$, $B^0_d \to D^+_d    \tau^- \bar{\nu}_{\tau}$ and $B^0_d \to D^{*+}_d \tau^- \bar{\nu}_{\tau}$ decays.
The precise measurements of these semileptonic branching ratios have been reported by BABAR, Belle and LHCb \cite{Lees:2012xj,Lees:2013uzd,Huschle:2015rga,Aaij:2015yra}.
The 95\% CL experimental ranges of the average data from PDG \cite{Agashe:2014kda} and the 95\% CL SM predictions are listed in the second and the third columns of Table~\ref{tab:bctaulslp}, respectively.
\begin{table}[b]
\caption{Branching ratios of the exclusive $ b\to c \tau^-\bar{\nu}_{\tau} $ decays (in units of $10^{-2}$). }
\begin{center}
\begin{tabular}{lcccc}
\hline\hline
 Observable                                              &Exp. data       & SM predictions    & SUSY/$\lambda_{i33}\tilde{\lambda}'^*_{i23}$    \\\hline
$\mathcal{B}(B^-_c\to        \tau^- \bar{\nu}_{\tau})$    &$\cdots$         &$[1.42,~2.78]$      &$[0.87,100]$     \\\hline
$\mathcal{B}(B^-_u\to D^0_u \tau^- \bar{\nu}_{\tau})$    &$[0.28,~1.26]$    &$[0.52,~0.90]$      &$[0.64,1.20]$    \\\hline
$\mathcal{B}(B^-_u\to D^{*0}_u \tau^- \bar{\nu}_{\tau})$ &$[1.49,~2.27]$    &$[1.21,~1.47]$      &$[1.21,1.53]$    \\\hline
$\mathcal{B}(B^0_d\to D^+_d \tau^- \bar{\nu}_{\tau})$    &$[0.60,~1.46]$    &$[0.48,~0.84]$      &$[0.60,1.12]$    \\\hline
$\mathcal{B}(B^0_d\to D^{*+}_d \tau^- \bar{\nu}_{\tau})$  &$[1.41,~2.27]$    &$[1.12,~1.36]$     &$[1.12,1.41]$    \\\hline
\hline
\end{tabular}
\end{center}
\label{tab:bctaulslp}
\end{table}

\begin{figure}[t]
\begin{center}
\includegraphics[scale=0.6]{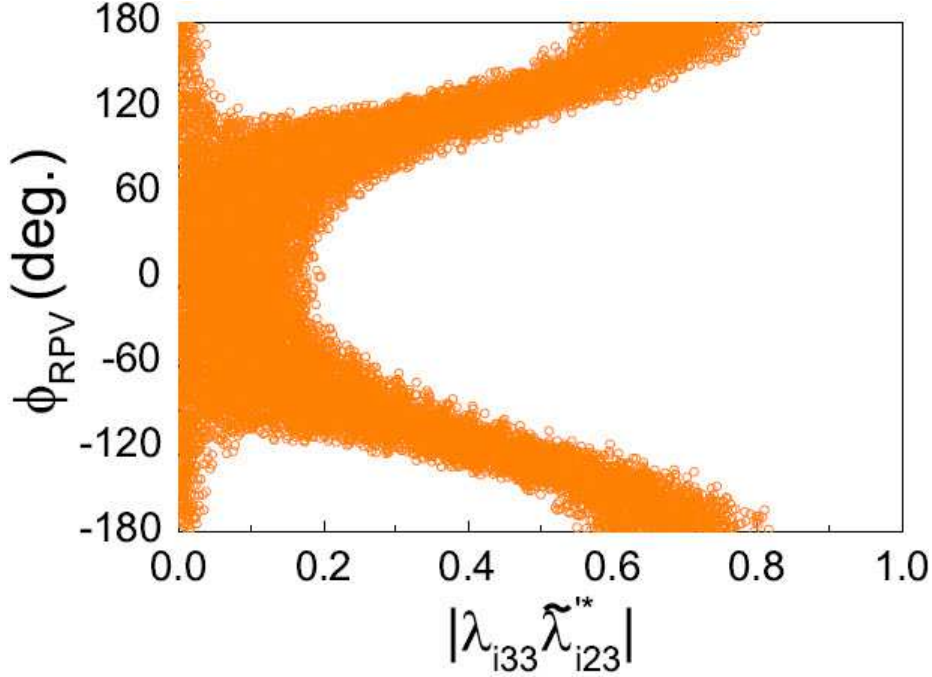}
\end{center}
 \caption{The allowed parameter spaces of $\lambda_{i33}\tilde{\lambda}'^*_{i23}$  from the  95$\%$ CL experimental bounds of   the exclusive $b\to c\ell^-\bar{\nu}_\ell$ decays. }
 \label{fig:RPVcouling}
\end{figure}

Fig.~\ref{fig:RPVcouling} displays the allowed parameter spaces of the couplings $\lambda_{i33}\tilde{\lambda}'^*_{i23}$  from the 95$\%$ CL experimental bounds of the exclusive $b\to c\ell^-\bar{\nu}_\ell$ decays.  Both the moduli and the weak phases of  $\lambda_{i33}\tilde{\lambda}'^*_{i23}$
are obviously constrained by current experimental measurement. The bounds on $\lambda_{i33}\tilde{\lambda}'^*_{i23}$ is obtained for the first time.
\begin{figure}[t]
\begin{center}
\includegraphics[scale=0.7]{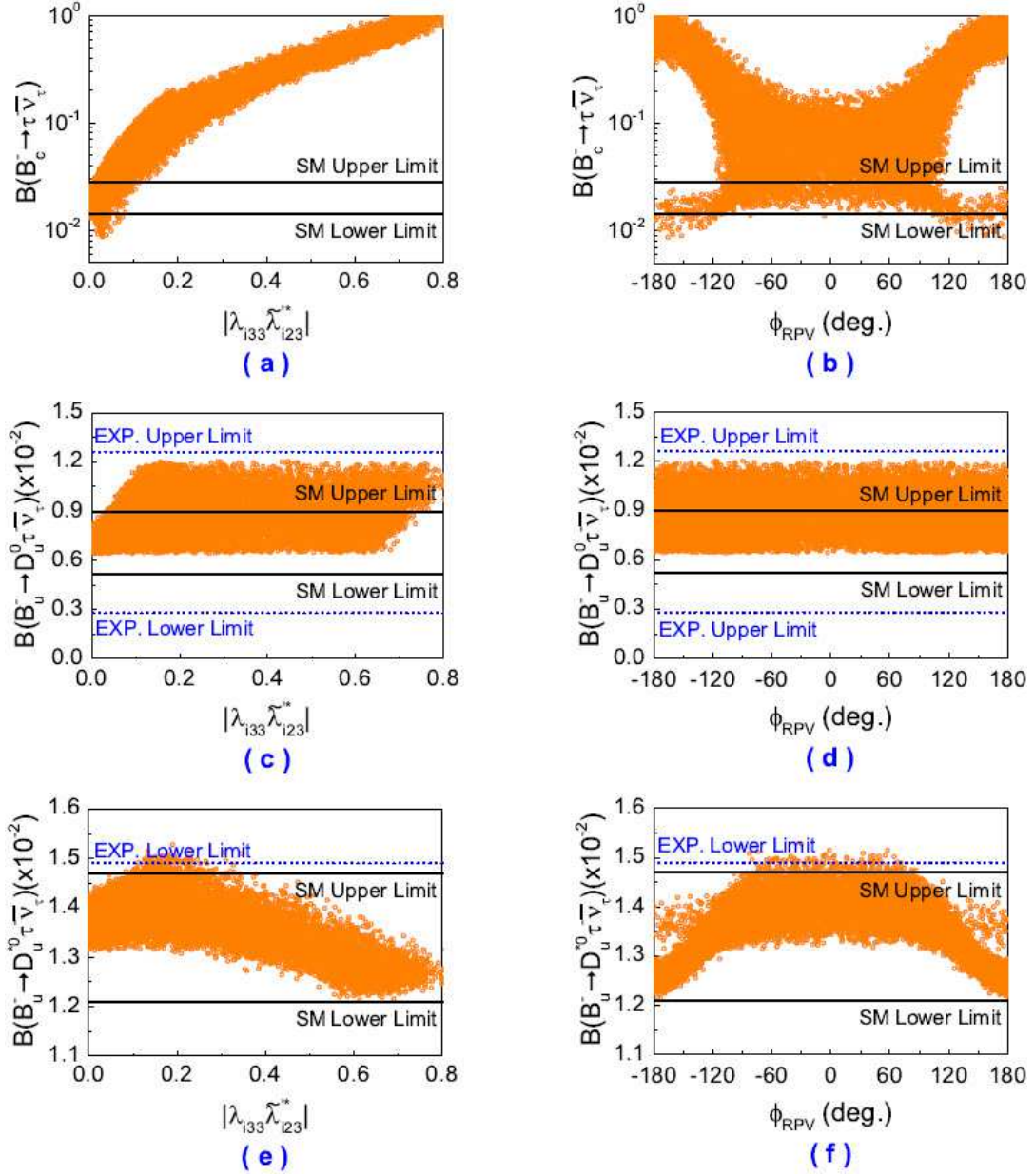}
\end{center}
 \caption{The constrained effects of the slepton exchange coupling $\lambda_{i33}\tilde{\lambda}'^*_{i23}$   in the  exclusive $b \to c \tau^- \bar{\nu}_{\tau}$ decays. }
 \label{fig:Brbctaulslp}
\end{figure}
\begin{figure}[t]
\begin{center}
\includegraphics[scale=0.7]{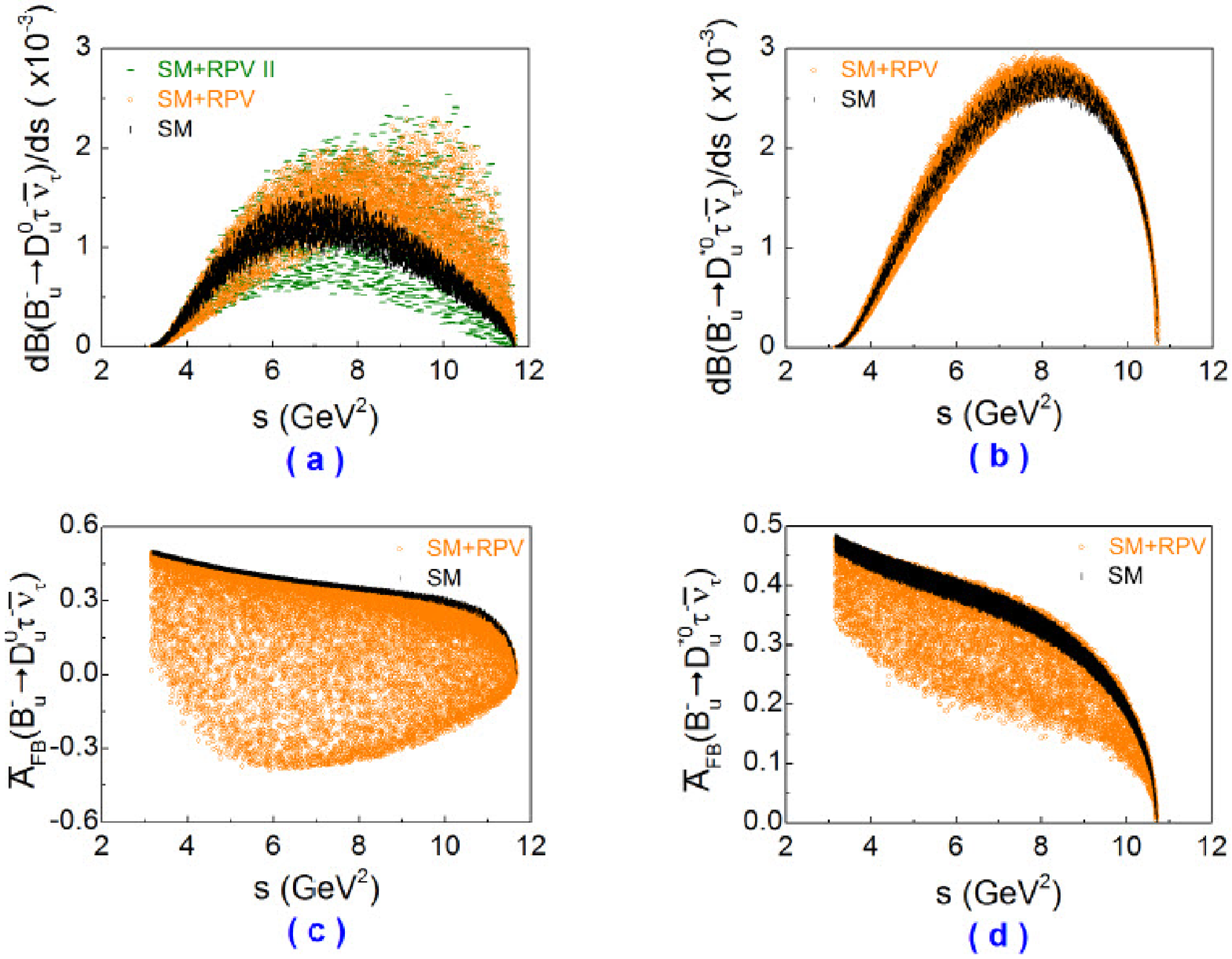}
\end{center}
 \caption{The constrained slepton exchange effects  on the differential branching ratios and the normalized FB asymmetries of $B^-_u\to D^{(*)0}_u \tau^- \bar{\nu}_{\tau}$ decays. }
 \label{fig:bctaulslp}
\end{figure}

Now we discuss  the constrained $\lambda_{i33}\tilde{\lambda}'^*_{i23}$  effects in the exclusive $ b\to c \tau^-\bar{\nu}_{\tau} $ decays. Our numerical predictions are given in the last column of Table~\ref{tab:bctaulslp}.  Fig.~\ref{fig:Brbctaulslp} shows the sensitivities of branching ratios to both moduli and weak phases of $\lambda_{i33}\tilde{\lambda}'^*_{i23}$, and  Fig.~\ref{fig:bctaulslp} shows the constrained slepton exchange effects  on the differential branching ratios and the normalized FB asymmetries of $B^-_u\to D^{(*)0}_u \tau^- \bar{\nu}_{\tau}$ decays.

As displayed in Fig.~\ref{fig:Brbctaulslp} (a-b), $\mathcal{B}(B^-_c \to \tau^- \bar{\nu}_\tau)$ is very sensitive to both moduli and weak phases of $\lambda_{i33}\tilde{\lambda}'^*_{i23}$, so the future  experimental measurements on $\mathcal{B}(B^-_c \to \tau^- \bar{\nu}_\tau)$ will give quite strong bound on  $\lambda_{i33}\tilde{\lambda}'^*_{i23}$.
 As displayed in Fig.~\ref{fig:Brbctaulslp} (c-d), $\mathcal{B}(B^-_u \to D^{0}_u\tau^- \bar{\nu}_{\tau})$ is sensitive to $|\lambda_{i33}\tilde{\lambda}'^*_{i23}|$ but  not very sensitive to their weak phases. We also can see that present experimental measurements of $\mathcal{R}(D)$ give strong bounds on this branching ratio.
As displayed in Fig.~\ref{fig:Brbctaulslp} (e-f), $\mathcal{B}(B^-_u \to D^{*0}_u\tau^- \bar{\nu}_{\tau})$ is very sensitive to both moduli and weak phases of $\lambda_{i33}\tilde{\lambda}'^*_{i23}$ couplings,  $\mathcal{B}(B^-_u \to D^{*0}_u\tau^- \bar{\nu}_{\tau})$ could have maximum at $|\lambda_{i33}\tilde{\lambda}'^*_{i23}| \in [0.1,0.2]$ and $\phi_{RPV} \in [-60^{\circ}, 60^{\circ}]$, and they could catch the lower limits of present 95\% CL experimental averages.

In Fig.~\ref{fig:bctaulslp} (a), we show another RPV prediction with the green ``-" labeled with ``SM+RPVII", which is  constrained by all  above mentioned 95\% CL experimental measurements except $\mathcal{R}(D)$.  We can see that the constrained $\lambda_{i33}\tilde{\lambda}'^*_{i23}$ couplings have very large effects on  $d\mathcal{B}(B^-_u \to D^0_u    \tau^- \bar{\nu}_{\tau})/ds$ at whole $s$ regions, and the 95\% CL experimental bound of  $\mathcal{R}(D)$ give quite obvious constraints at middle and high $s$ regions.
Fig.~\ref{fig:bctaulslp} (b) shows us that  the constrained $\lambda_{i33}\tilde{\lambda}'^*_{i23}$ couplings have some effects on $d\mathcal{B}(B^-_u \to D^{*0}_u\tau^- \bar{\nu}_{\tau} )/ds$ at the middle $s$ region.
As displayed in Fig.~\ref{fig:bctaulslp} (c-d), the constrained $\lambda_{i33}\tilde{\lambda}'^*_{i23}$ couplings have significant effects on $\bar{\mathcal{A}}_{FB}(B^-_u \to D^{(*)0}_u \tau^- \bar{\nu}_\tau)$ at whole $s$ region.

 \subsection{The ratios $\mathcal{R}(D)$ and $\mathcal{R}(D^*)$}

 For the exclusive  $b \to c \ell^- \bar{\nu}_{\ell}$ decays,  the ratios of the branching ratios have been accurately measured by LHCb, BABAR  and Belle\cite{Lees:2012xj, Lees:2013uzd,Huschle:2015rga,Aaij:2015yra}. At 95\% CL, their experimental averaged ranges, the SM predictions and the RPV SUSY predictions  are listed in Table~\ref{tab:ratios}. We can see that $\mathcal{R}(D)$ is constrained by its 95\% CL experimental measurements. As for $\mathcal{R}(D^*)$,
 the maximum of the RPV prediction is almost reach the lower limit of its 95\% CL experimental measurements.

\begin{figure}[t]
\begin{center}
\includegraphics[scale=0.65]{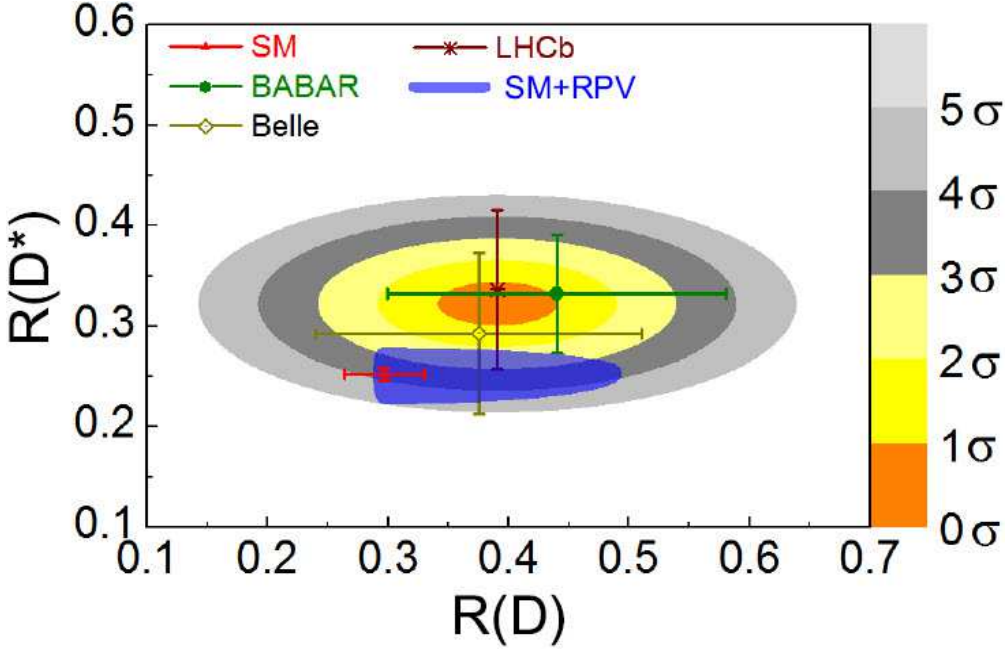}
\end{center}
 \caption{The ratios $\mathcal{R}(D)$ and $\mathcal{R}(D^{*})$. The theoretical predictions and experimental measurements from BABAR, Belle and LHCb are shown at 95\% CL, and the experiential average are given within $5\sigma$.  }
 \label{fig:b2carea}
\end{figure}

In order to compare easily, we display the 95\% CL SM predictions, the  95\% CL RPV SUSY predictions, the  95\% CL experimental measurements from BABAR, Belle as well as  LHCb,  and their experimental  average within $5\sigma$  on the $\mathcal{R}(D)$-$\mathcal{R}(D^*)$ plane in Fig.~\ref{fig:b2carea}, and we can clearly  see that  our RPV SUSY predictions have about 2$\sigma$ deviations from the experimental averaged values  on the $\mathcal{R}(D)$-$\mathcal{R}(D^*)$ plane.
 \begin{table}[t]
\caption{The ratios $\mathcal{R}(D)$ and $\mathcal{R}(D^{*})$ in the exclusive $ b\to c \ell^-\bar{\nu}_{\ell} $ decays.}
\begin{center}
\begin{tabular}{lccc}
\hline\hline
$\mathcal{R}(D)$      &$[0.294,~0.488]$                          &$[0.251,~0.343]$        &$[0.294,~0.488]$           \\\hline
$\mathcal{R}(D^*)$    &$[0.280,~0.364]$                          &$[0.242,~0.263]$         &$[0.226,~0.278]$          \\\hline
\hline
\end{tabular}
\end{center}
\label{tab:ratios}
\end{table}
\begin{figure}[t]
\begin{center}
\includegraphics[scale=0.7]{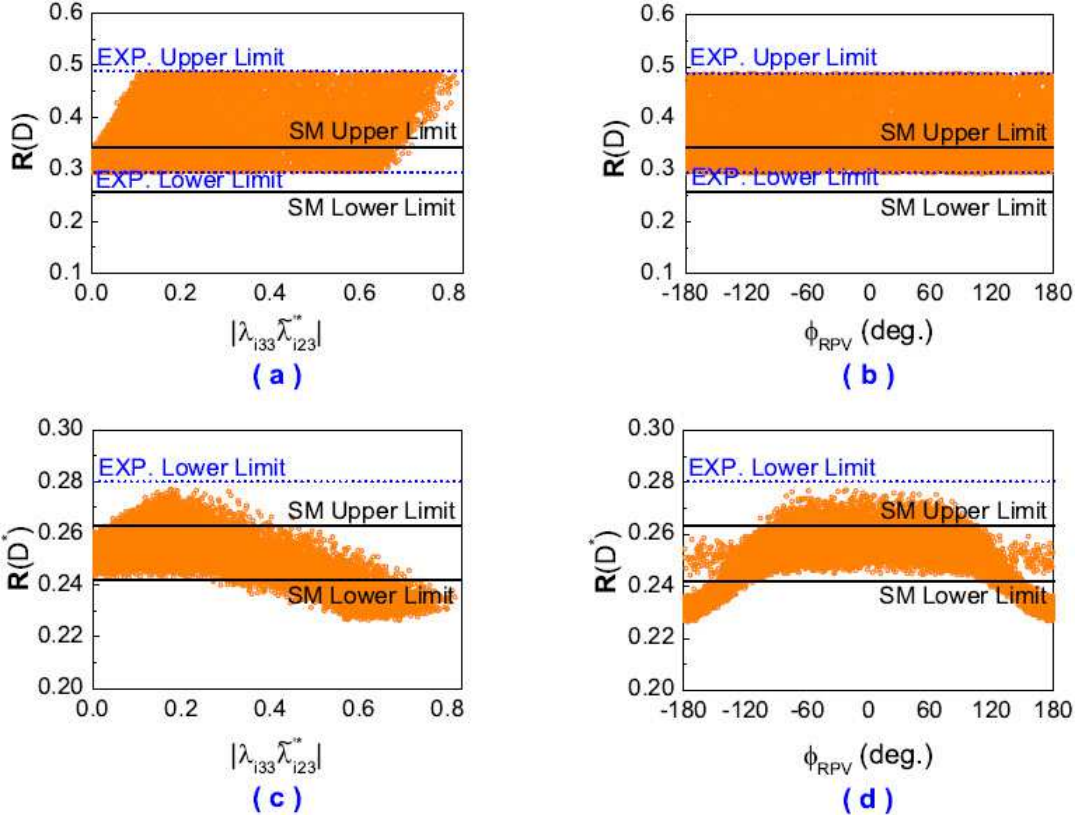}
\end{center}
 \caption{The constrained effects of RPV coupling $\lambda_{i33}\tilde{\lambda}'^*_{i23}$  due to the slepton exchange in the ratios $\mathcal{R}(D^{(*)})$. }
 \label{fig:RDlslp}
\end{figure}
At 95\% CL, the RPV SUSY predictions of $\mathcal{R}(D)$ and $\mathcal{R}(D^{*})$  are consistent with each experimental measurement from BABAR, Belle and LHCb. The error of the experiential average is much smaller than each one of measurements from BABAR, Belle and LHCb,  at 99\% CL, the RPV SUSY predictions for $\mathcal{R}(D)$ and $\mathcal{R}(D^{*})$ agree with the experimental  averages.

Now we give the sensitivities of  $\mathcal{R}(D^{(*)})$ to the slepton exchange couplings.
Since the ratios $\mathcal{R}(D)$ and $\mathcal{R}(D^*)$ are not sensitive to $\lambda_{i11}\tilde{\lambda}'^*_{i23}$ and $\lambda_{i22}\tilde{\lambda}'^*_{i23}$ couplings, we only show the sensitivities to the moduli and weak phases of  $\lambda_{i33}\tilde{\lambda}'^*_{i23}$ couplings in Fig.  \ref{fig:RDlslp}.
As displayed in Fig. \ref{fig:RDlslp} (a-b), $\mathcal{R}(D)$ is  sensitive to $|\lambda_{i33}\tilde{\lambda}'^*_{i23}|$ coupling, and the experimental  average of $\mathcal{R}(D)$ gives obvious constraints on $|\lambda_{i33}\tilde{\lambda}'^*_{i23}|$.
In Fig. \ref{fig:RDlslp} (c-d), $\mathcal{R}(D^*)$ is very sensitive to both moduli and weak phases of the $\lambda_{i33}\tilde{\lambda}'^*_{i23}$ couplings, it could have maximum at $|\lambda_{i33}\tilde{\lambda}'^*_{i23}| \in [0.1,0.2]$ and $\phi_{RPV} \in [-60^{\circ}, 60^{\circ}]$.

\section{Conclusion}
Motivated by the recent experimental data of the ratios $\mathcal{R}(D^{(*)})$ reported by LHCb, BABAR and Belle collaborations, we have studied the RPV SUSY effects in the leptonic and semileptonic decays, $B^-_c \to \ell^- \bar{\nu}_{\ell}$, $B^-_u \to D^{0}_u \ell^- \bar{\nu}_{\ell}$, $B^-_u \to D^{*0}_u \ell^- \bar{\nu}_{\ell}$, $B^0_d \to D^+_d \ell^- \bar{\nu}_{\ell}$, $B^0_d \to D^{*+}_d \ell^- \bar{\nu}_{\ell}$.
Considering the theoretical uncertainties   and the experimental
errors at $95\%$ CL, we have constrained  the parameter spaces of relevant RPV couplings from the present experimental data. We have found that  the effects of the squark exchange couplings could be neglect in the exclusive $b \to c \ell^- \bar{\nu}_{\ell}$  decays.  As for the slepton exchange couplings, the strongest bounds on $\lambda_{i11}\tilde{\lambda}'^*_{i23}$  and $\lambda_{i22}\tilde{\lambda}'^*_{i23}$  came from  the exclusive  $b\to s \ell^+ \ell^-$ decays, and the bounds on $\lambda_{i33}\tilde{\lambda}'^*_{i23}$ have been obtained from the exclusive $b\to c \ell^- \bar{\nu}_{\ell}$ decays for the first time.

Furthermore, we have predicted the constrained  slepton exchange  effects on the branching ratios, the differential branching ratios and  the normalized FB asymmetries of the charged leptons, the ratios of the semilepton decay branching ratios.
 We have found that $\mathcal{B}(B^-_c \to \ell^- \bar{\nu}_\ell)$ and $\mathcal{B}(B \to D^{*} \tau^- \bar{\nu}_{\tau})$  are very sensitive to the constrained slepton exchange couplings, and the constrained slepton exchange couplings have great effects on $d\mathcal{B}(B \to D \tau^- \bar{\nu}_{\tau})/ds$,  $\bar{\mathcal{A}}_{FB}(B \to D e^- \bar{\nu}_{e})$,   $\bar{\mathcal{A}}_{FB}(B \to D \tau^- \bar{\nu}_{\tau})$ and   $\bar{\mathcal{A}}_{FB}(B \to D^* \tau^- \bar{\nu}_{\tau})$,  in addition, the sign of $\bar{\mathcal{A}}_{FB}(B \to D e^- \bar{\nu}_{e})$ could be changed by the large slepton exchange couplings $\lambda_{i11}\tilde{\lambda}'^*_{i23}$.

 For  $\mathcal{R}(D)$ and  $\mathcal{R}(D^{*})$, they are very sensitive to the constrained $\lambda_{i33}\tilde{\lambda}'^*_{i23}$ couplings, but  not sensitive to  the constrained $\lambda_{i11}\tilde{\lambda}'^*_{i23}$ and $\lambda_{i22}\tilde{\lambda}'^*_{i23}$ couplings from the exclusive $b\to s \ell^+\ell^-$ decays.
 The constrained $\lambda_{i33}\tilde{\lambda}'^*_{i23}$ couplings could enhance $\mathcal{R}(D)$ to its 95\% CL experimental range.  Although the constrained $\lambda_{i33}\tilde{\lambda}'^*_{i23}$ couplings maybe enhance $\mathcal{R}(D^*)$, its maximum still   has 2$\sigma$ deviation from the 95\% CL experimental average.  Nevertheless, the constrained slepton exchange couplings could let $\mathcal{R}(D^*)$ to reach each 95\% CL experimental range from BABAR, Belle and LHCb.  In addition, at 99\% CL, the RPV prediction for $\mathcal{R}(D^*)$ agrees with the experimental  averages.

With the running LHCb and the forthcoming Belle-II experiments, heavy flavor physics is entering a precision era, which would present new features to examine various NP models, including the RPV SUSY model studied in this paper. Our results  could be useful for probing the RPV SUSY effects, and will correlate strongly
with searches for the direct RPV SUSY signals at future experiments.

\section*{Acknowledgments}
The work of Ru-Min Wang was supported by Program for New
Century Excellent Talents in University (No. NCET-12-0698).  The works of Yuan-Guo Xu, Ying-Ying Fan and Qin Chang were supported by the National Natural Science Foundation of China (Nos. U1204113, 11505148 and 11475055).

\end{document}